\documentstyle[12pt,epsfig]{article}
\begin{document}
\large
\begin{center}
{\Large On the production mechanism of $\Sigma$--hypernuclear systems
in A(K$^-, {\pi}^{\pm}$) reactions}\\[\baselineskip]
O.D.Dalkarov$^a$ and V.M.Kolybasov$^b$\\
{\it Lebedev Physical Institute, 117924 Moscow, Russia}
\end{center}
\vspace{0.5cm}
\baselineskip=24pt
{\bf Abstract}\\[\baselineskip]
\hspace*{0.4cm} It is shown that the new data on the excitation
energy $E_{\mbox{ex}}$ spectrum
of the residual nuclear system in the $\Sigma$--hypernuclear region in
the reactions $(\mbox{K}^-,{\pi}^{\pm})$ on $^9$Be and in the reaction
$(\mbox{K}^-,{\pi}^+)$ on $^4$He and $^{12}$C can be described without the
supposition on the existence of excited $\Sigma$--hypernuclear states.
The basis is formed by a simultaneous  consideration of the quasi--free
$\Sigma$--production
and $\Sigma$--nuclear rescattering (elastic and with
$\Sigma \to \Lambda$ conversion) with account of interference of the
respective amplitudes. For final decision of the question about the nature
of the irregularities in $E_{\mbox{ex}}$ spectrum, it is proposed to study
the picture corresponding to the so--called moving complex singularity of
the triangle graph with $\Sigma$ rescattering: the position and the width
of the peak in $E_{\mbox{ex}}$ distribution must appreciably change with
momentum transferred from the initial kaon to the final pion.
\\[\baselineskip]

\vspace{0.5cm}
\noindent -------------------------- \\
$^a$ e--mail: dalkarov@sci.lebedev.ru\\
$^b$ e--mail: kolybasv@sci.lebedev.ru

\newpage
\section{Introduction}

\hspace*{0.4cm} New BNL data on $^9\mbox{Be}(\mbox{K}^-,{\pi}^{\pm})$
reactions at 600 MeV/c in $\Sigma$--hypernuc\-lear region [1] drastically
changed the situation around the problem of the existence of excited
states of $\Sigma$--nuclei. As followed from the old data (see, for
instance, review [2]) there were clear indications on the narrow peaks
($\Gamma<10$ MeV) in the excitation energy spectrum of the residual
nuclear systems in the region close to $\Sigma$--hyperon production. For
this reason the idea about the creation of excited hypernuclear states
seemed to be very attractive. However, from the very beginning the
problem of a small hypernuclear width was discussed, since due to
$\Sigma N \to \Lambda N$ conversion in nuclear matter all estimations
lead to the widths more then $20 \div 40$ MeV [2,3]. Recent data (see
Fig.1 where circles correspond to $^9\mbox{Be}(\mbox{K}^-,{\pi}^-)$
reaction and squares to $^9\mbox{Be}(\mbox{K}^-,{\pi}^+)$ one) exclude
very narrow peaks but reveal the structures with the width about 20 MeV for
$(\mbox{K}^-, {\pi}^-)$ and $30 \div 40$ MeV for the case of
$(\mbox{K}^-, {\pi}^+)$ reaction.

Several questions should be cleared up: whether these peaks call for
the idea about $\Sigma$--hypernuclei existence or they are caused by
the reaction mechanism and, probably, by the nearthreshold phenomena?
If it is possible to understand the problem without $\Sigma$--hypernuclei
then how will the natural and doubtless description be made with the help
of the simplest mechanisms? Finally, are there the crucial tests to
clear the question about the nature of the irregularities in the
excitation energy spectrum of $\Sigma$--hypernuclear systems? Below we
will try to answer these questions.

The first goal of this study is to show that it is quite plausible to
describe all of the data on the reaction $^9\mbox{Be}(\mbox{K}^-,{\pi}^-)$
and the reactions $(\mbox{K}^-, {\pi}^+)$ on $^9$Be, $^{12}$C and $^4$He
nuclei without the idea
about the existence of excited $\Sigma$--hypernuclear states (see Fig.2d)
but using Feynman graph language and taking into account the quasi--free
$\Sigma$--hyperon production (Fig.2a), the elastic rescattering of $\Sigma$
(Fig.2b) and final inelastic interaction of $\Sigma$--hyperon with the
$\Sigma \to \Lambda$ conversion (Fig.2c). In this case the interference of
the pole graph of Fig.2a and the triangle graph of Fig.2b should be
essential. We will also emphasize some characteristic features of the
process $^9\mbox{Be}(\mbox{K}^-,{\pi}^+)$ distinguishing it against
others.

Another and the main purpose is to advance theoretical apparatus for final
revealing of the nature of the peaks in the excitation energy spectra
in order to give a method which could allow to distinguish the peaks caused
by the $\Sigma$--hypernuclei existence from ones produced by the reaction
mechanism. This method is based on the analytical properties of
Feynman graphs. In our case the singularities of the nonrelativistic
triangle graph (see Fig.2e) are close to the physical region. This fact
leads to the appearance of moving
maxima in the excitation energy spectra as a function of the square of 
the momentum transferred from initial kaon to final ${\pi}$-meson. Note that
in the experiment the extraction of this graph as a unique one is possible,
if the ${\Lambda}$-hyperon, produced by the interaction of virtual
${\Sigma}$-hyperon with the intermediate nucleus, is detected. Therefore,
the studying of the double differential cross sections (with and without 
${\Lambda}$-hyperon detection in the final state) for $A(K,{\pi})X$
reactions could be sufficient test to distinguish the main
features of the reaction mechanism.

The paper is organized as follows. The theoretical model is described
in Section 2. Kinematical relations between various differential cross
sections are given and the detailed properties of the amplitude for
the triangle graph are discussed. Section 3 is devoted to the procedure
of $^9$Be data processing, in particular, to a difference method which was
used to extract the contribution of $({\mbox{K}}^-,{\pi}^-)$ process on
the outer weekly bound neutron. The question about the role of relative
phase between the amplitudes for pole and triangle graphs (see Fig.2a
and 2b) is also discussed.

The final results for $({\mbox{K}}^-,{\pi}^-)$ reaction on $^9$Be and
$({\mbox{K}}^-,{\pi}^+)$ one on $^9$Be, $^4$He and $^{12}$C nuclei, which
are in good agreement with the experimental data, are given in Section 4.
We also discuss possible reasons for strong difference between the
excitation energy spectrum for $^9$Be and the same ones for $^4$He and
$^{12}$C.

The picture of the moving triangle singularities is discussed in Section
5. We present results of the calculations for the excitation energy
spectra for the channels with $\Sigma \to \Lambda$ conversion for
different momentum transfer from initial kaon to final pion. They show
that the moving peaks in the excitation energy spectra are experimentally
observable. For comparison the excitation energy spectra with hypernuclear
state production (see Fig.2d) are also calculated. In this case the
position of the peak does not practically depend on the momentum transfer.

The main results and concluding remarks are given in Conclusion.

\section{Theoretical model}

\hspace*{0.4cm} We will consider the graphs of Fig.2a--2c where, as
mentioned earlier, the pole graph (Fig.2a) represents quasi--free
$\Sigma$--hyperon production. Triangle graphs
correspond to the rescattering of virtual $\Sigma$ on the intermediate
nuclear system without conversion (Fig.2b) and with conversion (Fig.2c),
excluding production of $\Sigma$--hypernuclear bound or resonance states
(this process would correspond to the graph of Fig.2d). Let's analyze more
accurately the main properties of triangle Feynman graphs before to make
the fitting procedure. We will consider the general form of Fig.2e implying
that the particle 2 is $\Sigma$--hyperon, and the particle 1 is the residual
nuclear system and the lower vertex stands in principle for the aggregate
of all the processes that occur when $\Sigma$ interacts with the residual
nucleus. We denote by $p_i$ and $E_i$ the momentum and total energy of
a particle $i$ in the lab. system and introduce the notations
\begin{eqnarray}
&&{\bf q}={\bf p}_x-{\bf p}_z\, , \nonumber \\
&&\phantom{1} \\
&&W = \sqrt{s_{12}} = [(m_A+E_x-E_z)^2-q^2]^{1/2}\, .  \nonumber
\end{eqnarray}
Here $W$  is the invariant mass of the system {4 + \dots}n, consisting
of the particles  produced  after $\Sigma$--conversion
in nuclear medium.  We shall henceforth have relatively small
$q^2$ only and $W$ in the region where
the $\Sigma$--hyperon 2 can be assumed with good accuracy to  be
nonrelativistic.  If we neglect the complications that can appear when
account is taken of the spin structure of the amplitudes and  restrict
ourself to the consideration of triangle diagram only (Fig.2e) then
the quantity $ d^2 \sigma /dW\,dq^2 $ can be expressed in  terms  of
the  differential cross  section  $d {\sigma}_{3x}/d\Omega$
of the elementary reaction K$^{-}+$ N $ \to \pi+ \Sigma$ (in c.m.s.
of this reaction) and the total cross section ${\sigma}_{12}(W)$ for
the interaction of the $\Sigma$--hyperon and the nucleus 1 [4,5]:
\begin{equation}
\frac{d^2\sigma}{dW\,dq^2}  = \frac{m_1^2m_2s_{3x}}{4{\pi}^3m_3
(m_1+m_2)^2p_x^2}\kappa {\gamma}^2
\left(\frac{p_1^{cms}}{m_{12}}{\sigma}_{12}(W) \right)
\left( \frac{{\tilde p}_x}{{\tilde p}_z} \frac{d{\sigma}_{3x}}{d\Omega}
\right)|M|^2
\end{equation}
Here $s_{3x} = m_3^2 + m_x^2 + 2m_3 E_x \, ,\quad p_1^{cms}$
is the momentum of the relative  motion
of the particles 1 and 2 in the c.m.s. of particles 4\dots n,
${\tilde p}_x$ and ${\tilde p}_z$
are the momenta of particles x and z in the c.m.s. of the reaction
\mbox{3 + x $\to$ 2 + z}. The quantities ${\gamma}^2$ and $\kappa$
pertain to the nuclear vertex
\begin{equation}
           A \to 1 + 3 \, ,
\end{equation}
${\gamma}^2$ is the  reduced vertex part [6] and determines the probability of
the virtual disintegration (3), while $\kappa = \sqrt{2m_{13}\varepsilon}
\, ,\quad \varepsilon = m_1+m_3-m_A$. The factor $M$ is determined by the
structure of the triangle graph.

We will also need the differential cross section $d^2{\sigma}/d{\Omega}
dW$ for comparison with experimental data. It can be obtained from
Eq.(2) in the following manner:
\begin{equation}
\frac{d^2{\sigma}}{d{\Omega}\, dW} = \frac{W p_x p_z^2}{2{\pi}E_0
|p_z-\frac{E_z}{E_0} p_x \cos {\theta}|} \, \frac{d^2{\sigma}}{dW\, dq^2}
\end{equation}
where $\theta$ is the angle between particles $z$ and $x$ and $E_0$ is
the total energy of all particles in lab. system.

We shall henceforth focus our attention on the quantity $M$ which
determines the behaviour of differential cross section (2) as function
of kinematical variables (1). It is convenient to introduce
dimensionless variables [7]
\begin{eqnarray}
\xi & = & \frac{m_2}{m_3}\,\frac{m_A}{m_4+\cdots+m_n}\,
\frac{W-m_1-m_2}{\varepsilon} \nonumber \\
\phantom{1}\\
\lambda & = & \frac{m_1^2}{(m_1+m_2)^2}\, \frac{q^2}{{\kappa}^2} \nonumber
\end{eqnarray}
In terms of these variables $M$ can be expressed in the form of two--fold
integral in momentum space
\begin{equation}
M=\frac{1}{\kappa}\,\int\limits_0^{\infty}\int\limits_{-1}^1
\frac{F_l({\kappa}x)\,x^2dx\,P_l(z)\,dz}{(1+x^2)(x^2+\lambda-\xi
-2x\sqrt{\lambda}z-i{\eta})}
\end{equation}
with $x = p/{\kappa}$. Here $F_l(p)$ is the form factor of the vertex
A $\to 1+3$, normalized by the condition $F_l(i\kappa)=1$, $l$ is the
angular momentum of the relative motion of particles 1 and 3 in the
nucleus A, $P_l$ is the Legendre polynomial.

     In practice it is needed to use the general formulas
taking into account the realistic nuclear form factor.  In this case
it is simpler to make transformation to the coordinate space where
$M$ can be expressed as one--fold integral [4]
\begin{equation}
M= \frac{i^l}{2\pi}\, \int\limits_0^{\infty} {\Psi}(r)\,
j_l(\sqrt{\lambda}{\kappa}r) \, \exp(-A{\kappa}r+iB{\kappa}r)\,r\,dr \, .
\end{equation}
Here $j_l$ is the spherical Bessel function, the quantity ${\Psi}(r)$ is
introduced by the equation
\begin{equation}
{\Psi}(r)=4{\pi}i^l\,\int\limits_0^{\infty} \frac{F_l(p)}{p^2+{\kappa}^2}
\, j_l(pr)\,p^2\,dp
\end{equation}
(in the single-particle model it would be proportional to the wave
function of the relative  motion  of particles 1 and 3). The quantities
$A$ and $B$ are defined in the following manner
\begin{eqnarray}
&&B=\sqrt{\xi}\,,\quad A=0 \qquad \mbox{at} \quad {\xi}\ge 0 \nonumber \\
\phantom{a}\\
&&A=\sqrt{-\xi}\,,\quad B=0 \qquad \mbox{at} \quad {\xi}< 0 \, , \nonumber
\end{eqnarray}

Hereinafter, except Section 5, we will take the amplitude of the
lower vertex of Fig.2e to be constant as we are first of all interested
in the effects due to the structure and analytical properties of
the graphs. We would like, whenever possible, to obtain model
independent results. Taking into account that there are no reliable
data on sigma--nuclear interactions, we prefer not to rely on the
calculations using a $\Sigma$--A optical potential. Let us point to
the detailed research of K$^-$--$^4$He interactions with various kinds
of such potential [8]. In particular it shows a strong dependence
of the results on the potential parameters.

The amplitude $M$ (6) of Fig.2e graph has two types of singularities
in $W$: (i) normal threshold at $W=m_1 +m_2$,
and (ii) so--called triangle singularity
of logarithmic type which is situated in complex plane. The position
of the triangle singularity depends on the value of $q^2$ .
In terms of the variables $\xi$ and $\lambda$ the triangle singularity
is situated at
\begin{equation}
{\xi}_{\triangle} = \lambda - 1 + 2i\sqrt{\lambda} \, .
\end{equation}
If we can
approach closely to the position of the triangle singularity in an
experimental
investigation then the amplitude of a triangle graph would be
a sharp function and it is possible to expect that a bump in $W$ distribution
will appear. The position and width of the bump must vary with $q^2$. We
will discuss in Section 5 how this property of a triangle graph can be
checked.

\section{Procedure}

\hspace*{0.4cm} Though the data [1] on the processes
\begin{equation}
^9\mbox{Be}(\mbox{K}^-,{\pi}^+)
\end{equation}
and
\begin{equation}
^9\mbox{Be}(\mbox{K}^-,{\pi}^-)
\end{equation}
(see Fig.1) do not show any narrow structures, they, as we shall see
below, contain a lot of a physical information and unexpected features
(positions of bump maxima, an absence of narrow nearthreshold peaks due to
channels with $\Sigma \to \Lambda$ conversion and so on). The channel (11)
is related to $\Sigma$--production on the protons
\begin{equation}
\mbox{K}^- \mbox{p} \to {\pi}^+ {\Sigma}^-
\end{equation}
and the channel (12) can be realized on the protons
\begin{equation}
\mbox{K}^- \mbox{p} \to {\pi}^- {\Sigma}^+
\end{equation}
as well as on the neutrons
\begin{equation}
\mbox{K}^- \mbox{n} \to {\pi}^- {\Sigma}^0 \, .
\end{equation}
The cross section of the process (14) is much less than the cross section
of the process (15) at 600 MeV/c [2]. The data for the channels (11) and (12)
are quite different. The main reason, evidently, is very small binding
energy of the outer neutron in $^9$Be which is equal only 1.67 MeV. So
it is interesting to isolate the part of the channel (12) cross section
which takes place on the outer neutron.

Zero in the excitation energy $E_{\mbox{ex}}$ in the channel (11)
corresponds to the invariant mass $W$ of the final state consisting from
${\Sigma}^-$ plus the ground state of $^8$Li without relative motion
and in the channel (12) it corresponds to ${\Sigma}^0$ plus the
ground state of $^8$Be. So $E_{\mbox{ex}} \ge 0$ for events with
$\Sigma$--hyperon in a final state. Left parts of the spectra in
Fig.1, related to $E_{\mbox{ex}} < 0$, can have their origin in the
process of $\Sigma$ production followed by the conversion
\begin{equation}
\Sigma \mbox{N} \to \Lambda \mbox{N}
\end{equation}
as well as (for the channel (12)) in the ``tail'' of direct $\Lambda$
production. The estimation of this tail behaviour in the model of
a quasi--free $\Lambda$ production shows its sharp decrease in the
interval of $E_{\mbox{ex}}$ from -20 MeV to zero. It contradicts the
data on the channel (12). Therefore we take the model of a quasi--free
$\Lambda$ production followed by its rescattering. It leads to the
result shown by the solid curve in Fig.1 (the normalization of the curve
is fixed by the experimental point at $E_{\mbox{ex}}=-20$ MeV). In the
following the corresponding values (the physical background due to the
direct $\Lambda$ production) will be subtracted from the data for the
channel (12).

The nucleus $^9$Be has most probably a cluster structure which consists of
the core ($^8$Be or two $\alpha$--particles) and the loosely bound outer
neutron. So the reaction (12) can have a contribution from four protons and
four neutrons of the core as well as from the outer neutron. The reaction
(11) can proceed only on four core protons. We have simultaneously the data
on both channels (11) and (12). It provides a possibility to isolate the
part of the cross section of the channel (12) which is related to the
outer neutron contribution considering that the wave functions of the core
neutrons and protons are close. For this purpose let us note that the sum
of the cross sections of the processes (14) and (15) at 600 MeV/c is
approximately equal to 90 \% of the cross section of the process (13). Thus
we can believe that the contribution of the core neutrons and protons to
the cross section of the channel (12) is approximately 90 \% of the cross
section of the channel (11). Then the expression $({\sigma}_2 - 0.9
{\sigma}_1)$ gives the contribution of the outer neutron to the cross
section of the process (12). Here ${\sigma}_2$ is the cross section of the
channel (12) minus the contribution of the tail from the direct
$\Lambda$ production. In the following we will compare the results of our
calculations of the process (12) with the result of this very difference
procedure (see the points in Fig.4a).

In subsequent calculations we will use the wave function (form factor) of
the outer neutron in $^9$Be from the n--$\alpha$--$\alpha$ cluster model
[9]. The corresponding form factor for the core proton was not calculated in
the cluster model of ref.[9]. At the first stage we will use the p--wave
oscillator wave function with the parameter $p_0=130$ MeV/c [10].
For studying a sensitivity of the results to a shape of a wave function
we will also make calculations with the model p--wave function of a
``quasi--Hulten'' type
\begin{equation}
{\psi }(r) \sim \left(\frac{1}{{\kappa}r} + \frac{1}{{\kappa}^2r^2} \right) \left( e^{-{\kappa}r}
-3e^{-({\kappa}+{\rho})r}+ 3e^{-({\kappa}+2{\rho})r}- e^{-({\kappa}+3{\rho})r} \right) \,
\end{equation}
which has correct asymptotic behaviour at $r \to 0$ and $r \to \infty$. Note
at once that it will not change the results qualitatively.

Let us present at first several intermediate
results for the case of the reaction (11) at 600 MeV/c for small angles.
Fig.3a shows the real and imaginary parts of the triangle graph with a
secondary interaction of the $\Sigma$--hyperon with the residual nuclear
system (Fig.2b and 2c) as functions of $E_{\mbox{ex}}$. As was earlier
mentioned, the calculations were carried out with constant amplitude of
a secondary interaction in order to clear up at the first place what is
given by the structure of the graphs. Fig. 3b demonstrates the modulus
squared of the triangle graph amplitude. We can see that it has a sharp
peak near $E_{\mbox{ex}}=0$ with the width about 15 MeV. The cross section
of the graph of Fig.2b process includes the phase space factor proportional
to $\sqrt{E_{\mbox{ex}}}$ and it leads to the smoothing and shifting of the
peak. It is not the case for the process with the conversion (16) and
the corresponding peak must be present also in its cross section. Note
that the peak of the same nature is well known for the process
$({\mbox{K}}^-,{\pi}^-)$ on the deuteron [11]. The cusp structures are
also distinctly seen in the results of calculations of stopped and
in--flight K$^-$ interactions with He$^4$ [8,12].

The solid curve of Fig.3c shows the shape of $E_{\mbox{ex}}$ distribution
corresponding to the quasi--free $\Sigma$ production (the pole graph of
Fig.2a). The dotted curve shows the same for the triangle graph of
Fig.2b. We see that both the pole graph and the triangle graph separately
lead to the bumps with the width $30 \div 40$ MeV but with maxima in the
region of 10 MeV and it contradicts the experimental data. However, the
amplitudes of the graphs of Fig.2a and 2b interfere with each other.
Comparison of the real and imaginary parts of the triangle graph in Fig.3a
indicates that its phase varies sharply with $E_{\mbox{ex}}$ and the result
of above mentioned interference must be nontrivial. The dashed and
dash--dotted curves in Fig.3c are the results of calculations for the sum
of the graphs of Fig.2a and 2b with the relative phase equal to $0.4\pi$
and $0.9\pi$ respectively. They show that the position and the shape of
the resulting peak may be varied over a wide range by means of relative
phase variation. (Note that this phase is not known a priori as it is
essentially determined by the phase of the elastic $\Sigma$--nucleus
scattering amplitude and by the possible energy variation of the phases
of the elementary processes (13)--(15)). Large interference effects were
also noted in ref. [8].

\section{Results}

\hspace*{0.5cm} Let us pass to the presentation of the results for the
best fit to the data on the channels (11) and (12). First of all we are
interested in a principal possibility of the description without the
introduction of $\Sigma$--nuclei. Therefore at this point we did not set
a task
to estimate the absolute values of cross sections (at least it demands
to account additionally for the absorption in initial and final states) but
were concentrated on the description of the shape of the $E_{\mbox{ex}}$
distributions at small pion angles. For this reason the normalization
factors of Fig.2a and 2b graphs and their relative phase were taken
as free parameters. Here it is necessary to make a few notes. As to an
absolute normalization, ref. [8] shows that the theoretical calculations
for $^4$He case lead to the reasonable results when accounting of kaon and
pion waves absorption. Relative contribution and phase of Fig.2b graph
now cannot be evaluated reliably due to a lack of information on
sigma--nuclear interactions. It is possible to put the inverse task
about deriving an information on sigma--nuclear interaction from
outcomes of a comparison of calculations with experimental data. It,
however, is a theme of an independent research.

The solid curve of Fig.4a shows the result of the calculation for the sum
of Fig.2a and 2b graphs with the relative phase $1.3\pi$ for the reaction
$^9\mbox{Be}(\mbox{K}^-,{\pi}^+)$. It agrees with the data very well.
The dashed curve corresponds to the "switching--off"
the triangle graph of Fig.2b, i.e. presents the separate contribution of
the quasi--free $\Sigma$ production (Fig.2a). The contribution of Fig.2c
graph was not taken into account as the experimental points at
$E_{\mbox{ex}}<0$ are practically equal to zero.

Fig.4b deals with the difference data for the reaction
$^9\mbox{Be}(\mbox{K}^-,{\pi}^-)$ which have their origin in the process
on the outer neutron (see the preceding section). The dotted curve
shows the supposed contribution from the process with the conversion
$\Sigma \to \Lambda$ (Fig.2c). Essentially it is analogue of the curve of
Fig.3b normalized to the point at $E_{\mbox{ex}}=0$ where the contributions
of Fig.2a and 2b processes go to zero. The solid curve is the result
of a full calculation with account of the interference of Fig.2a and 2b
graphs with the relative phase $1.9\pi$. The dashed curve is the separate
contribution of the quasi--free process. We notice that the cross section
of the process (12) on the outer neutron has the appearance of the peak in
$E_{\mbox{ex}}$ with a maximum in the region of $8 \div 10$ MeV, the
width of $15 \div 20$ MeV, and can be very well described by the
combination of Fig.2a--2c graphs.

Having obtained the good results for the production of
$\Sigma$--hypernuclear systems on $^9$Be, we pass now to the description
by the same method of the new data on the reaction $^4\mbox{He}
({\mbox{K}}^-,{\pi}^+)$  at 600 MeV/c [13]\footnote{We will not consider
now the process $^4\mbox{He}({\mbox{K}}^-,{\pi}^-)$  where the bound
$\Sigma$--hypernuclear state of $^4$He was discovered [13]. Here the picture
is more complicated due to presence of a resonance peak. In principle,
our model must describe the background including, in particular, all
data at $E_{\mbox{ex}}>0$. We hope to discuss it in another publication.}
and on the reaction
$^{12}\mbox{C}({\mbox{K}}^-,{\pi}^+)$  at 715 MeV/c [14]. For $^4$He we
use the s--wave oscillator wave function with the parameter $p_0=
90$ MeV/c which gives the best fit to the data on the process
$^4$He(e,ep)$^3$H [15]. For $^{12}$C we use the p--wave oscillator wave
function with the parameter $p_0=80$ MeV/c which gives the best fit to
the data on the reaction $^{12}$C(e,ep) in the ground and low lying
states of $^{11}$B [16]. The results are shown in Fig.5a for $^{12}$C and
in Fig.5b for $^4$He. Here the dotted curves are the contributions of
the processes with the conversion normalized to the points at
$E_{\mbox{ex}}=0$. The solid curves present the results of full
calculations with account of the interference of Fig.2a and 2b graphs
with the relative phase equal to $0.9\pi$ for Fig.5a and $1.3\pi$
for Fig.5b. The dashed curves are the separate contributions of the
quasi--free $\Sigma$ production. One can see that our simple
model provides a possibility to describe the data very well.

Certainly, owing to use of large number of fitting parameters our
description of the data on (K$^-,{\pi}^{\pm}$) reactions can be considered
simply as a successful parametrization. However, the possibility of such
parametrization was not obvious beforehand. We shall note that in
sigma--nuclear physics the use of large number of free parameters is not
the unusual fact. Let us indicate, for example, the paper [12] where
four parameters were used for the description of stopped K$^-$ interaction
with $^4$He.

It is necessary to emphasize also the following. In our calculations it was
supposed that the residual nuclear system is in the ground state or in one
of low excited states. There is direct experimental data on the reaction
(e,ep) for the cases of $^4$He and $^{12}$C. They indicate that the
vertices of virtual break--up of these nuclei to proton and ground states
of t and $^{11}$B give dominant contribution [17,18]. The same is also
noted for $^{12}$C case in ref. [19] devoted to the quasifree $\Sigma$
production in $(\mbox{K}^-,{\pi}^+)$ reactions. There are no electron data
on the vertex $^9\mbox{Be} \to \mbox{n} + ^8\mbox{Be}$. However the
evaluation in $(2 \alpha + \mbox{n})$ model [9] shows a preference of the
transition to the ground state of $^8$Be. Apparently, it is not so for
the process $^9\mbox{Be}(\mbox{K}^-,{\pi}^+)$. This case will be
separately considered in the following section.

\section{The case of $^9\mbox{Be}(\mbox{K}^-,{\pi}^+)$ reaction}
\hspace*{0.5cm} $E_{\mbox{ex}}$ distribution for
$^9\mbox{Be}(\mbox{K}^-,{\pi}^+)$ reaction sharply differs from other
cases in two aspects: (i) its maximum is shifted rightwards to
$25\div 30$ MeV, whereas in all remaining cases it is located near 10
MeV; (ii) it contains very few events at $E_{\mbox{ex}}\le 0$ and
practically does not leave a place for the contribution of the narrow
nearthreshold peak of Fig.3b. So the description, presented in Fig.4a,
was obtained without the contribution of the conversion process and used
a certainly too large value for the oscillator parameter $p_0=130$
MeV/c. On the other hand, proceeding from known values of the cross
sections of the elementary processes ${\sigma}({\Sigma}^-\mbox{p} \to
{\Sigma}^-\mbox{p})$ and ${\sigma}({\Sigma}^-\mbox{p} \to {\Lambda}
\mbox{n})$, each of which is about 150 mb in $\Sigma$ momentum region
$100 \div 200$ MeV/c, it is possible to evaluate that the cross sections
of the processes of conversion and elastic ${\Sigma}^-$ rescattering
should be of the same order as the cross section of the quasi--free
production.

The situation with the contribution of the conversion process could be
explained by supposition that the secondary $\Sigma$ interactions in
the lower vertices of Fig.2b and 2c graphs proceed mainly in p--wave.
It results in smoothening of the nearthreshold peak and shifts it to
the right [20]. However this explanation does not seem natural as there
are no reasons for the special behaviour just in the
$^9\mbox{Be}(\mbox{K}^-,{\pi}^+)$ case. More logical is other
explanation. It is probable that continuum states of the residual nuclear
system $^8$Li dominate in the break--up vertex $^9\mbox{Be} \to
\mbox{p} + ^8\mbox{Li}$. Contrary to the $^4$He and $^{12}$C cases,
there are no high resolution data on $^9$Be(e,ep) reaction. It is
possible only to state that the available data [21] show a wide
distribution with respect to the proton removal energy and do not
contradict such hypothesis. In that case, on the one hand, the
$E_{\mbox{ex}}$ distribution from the quasi--free $\Sigma$ production
is shifted to the right. On the other hand, the intermediate state in
Fig.2c graph becomes not two--particle but three-- or many--particle.
It completely changes the shape of nearthreshold behaviour. Fig.6
shows the comparison of $|M|^2$ for the triangle graphs with
two--particle (solid curve) and three--particle (dotted curve)
intermediate states. The character of the dotted curve leaves room for
the significant contribution of the conversion, keeping small number of
events at $E_{\mbox{ex}}\le 0$. Fig.7 shows an example of successful
description of $^9\mbox{Be}(\mbox{K}^-,{\pi}^+)$ data with considerable
contribution of the conversion process (dotted curve). Here the dashed
curve is the separate contribution of the quasi--free $\Sigma$ production.
Solid curve is the summary result with account of Fig.2a and 2b graphs
interference, relative phase being 0.35$\pi$. The value of the
oscillator parameter $p_0=115$ MeV/c was used. It is close to the value
110 MeV/c which is suggested in ref. [21] for the region of large
proton removal energies.

\section{Moving singularities and the mechanism of $\Sigma$--hypernuclear
systems production}

\hspace*{0.5cm} Strictly speaking, the good description of the data on the
$(\mbox{K}^-,{\pi}^{\pm})$ processes in the $\Sigma$--hypernuclear region by
the simplest mechanisms does not exclude a possible contribution from
$\Sigma$--hypernuclei. For a complete and unambiguous solution of the
reaction mechanism problem it seems efficient to use the theoretical
predictions which follow from the picture of moving complex triangle
singularities described in Section 2. As mentioned above, the presence of
these singularities  near the physical region of a reaction must lead to
a maximum in $E_{\mbox{ex}}$ distribution. The position and the shape of
the bump must change with the momentum $q$ transferred from the initial
kaon to the final pion [5]. Numerical calculations should show whether this
effect is noticeable or not. The contribution of the quasi--free
$\Sigma$--hyperon production would conceal the above mentioned effect.
Therefore it is desirable to study it in the channels with the conversion
$\Sigma \to \Lambda$ (i.e. with the detection of $\Lambda$) where
Fig.2a graph does not make a contribution. To investigate the discussed
picture one needs to measure the differential cross section $d^2{\sigma}/
dE_{\mbox{ex}}dq^2$, which is directly expressed through the modulus
squared of the matrix element (see Eq.(2)), in a wide range of
$E_{\mbox{ex}}$ and $q$.

For example, let us consider the reaction $^{12}\mbox{C}(\mbox{K}^-,{\pi}^+)$.
Fig.8a shows the theoretical predictions for the modulus squared of the
Fig.2c graph amplitude as a function of $E_{\mbox{ex}}$ for different values
of q= 200, 250, 300, 350 MeV/c. A distinct picture of the moving and the
broadening of the peak is visible. This picture is quite available for an
experimental observation.

The question is what would happen with the same distributions in the case of
$\Sigma$--hypernucleus production (the graph of Fig.2d)? To answer the
question,
calculations were made with inclusion of a resonance state (a Breit--Wigner
pole was put in) with the width of 10 MeV and the mass which was 15 MeV
more the sum of the masses of $\Sigma$ and the ground state of residual
nucleus. Fig.8b shows the results for the same set of the momentum transfer.
As could be expected, the position of the maximum remains practically
constant in this case. It follows that the investigation of  $d^2{\sigma}/
dE_{\mbox{ex}}dq^2$ would make it possible to answer unambiguously  the
question about the nature of the irregularities in the excitation energy
spectrum of the processes $({\mbox{K}}^-, {\pi}^{\pm})$: whether they are
due to the reaction mechanism or to the existence of $\Sigma$--hypernuclei.

\section{Conclusion}
\hspace*{0.5cm} Thus all considered data on the reactions
$({\mbox{K}}^-, {\pi}^{\pm})$ in the $\Sigma$--hypernuclear region  can be
basically described without the supposition on the existence of
$\Sigma$--hypernuclei. The bumps in the excitation energy distributions of
the residual nuclear systems are due to the peculiarities of the reaction
mechanisms.

The successful description of the available data by means of the simplest
mechanisms cannot completely exclude the existence of hypernuclei. We
tried to emphasize that the decisive conclusion on this problem can be
made only with the help of a detailed investigation of the
$\Sigma$--hypernuclear system production mechanism. We propose to study
the cross section $d^2{\sigma}/dE_{\mbox{ex}}dq^2$ at different values of
momentum transfer $q$, as its behaviour strongly depends on the existence
of $\Sigma$--hypernuclei. If the experimental investigations will confirm
the picture of moving singularities, predicted in Section 5, and thus the
dominant contribution of the Fig.3c graph in the channels with the
conversion, then it would be possible to extract the cross section
${\sigma}_{12}$ of the $\Sigma$--nucleus interaction using Eq.(2).
In the future this value of ${\sigma}_{12}$ could be compared with
dynamical calculations.

Note that the considered picture of moving singularities of triangle
Feynman graphs in the case of rescattering effects for
$\Sigma$--hypernuclear systems production is universal one. The same
phenomena could be observed at different kinematical conditions in other
reactions, for instance A(e, e$'$K$^-$)X.

The authors are indebted to Prof. I.S.Shapiro for the attention to the
investigation and discussions, to Profs. T.Nagae and R.E.Chrien for
information on the experimental data and to Profs. V.I.Kukulin,
V.I.Pomerantsev and M.A.Zhusupov for the advices and the data on the
neutron form factor of $^9$Be.

\newpage
\begin{figure}
\begin{picture}(500,300)
\put(20,-10){\epsfig{file=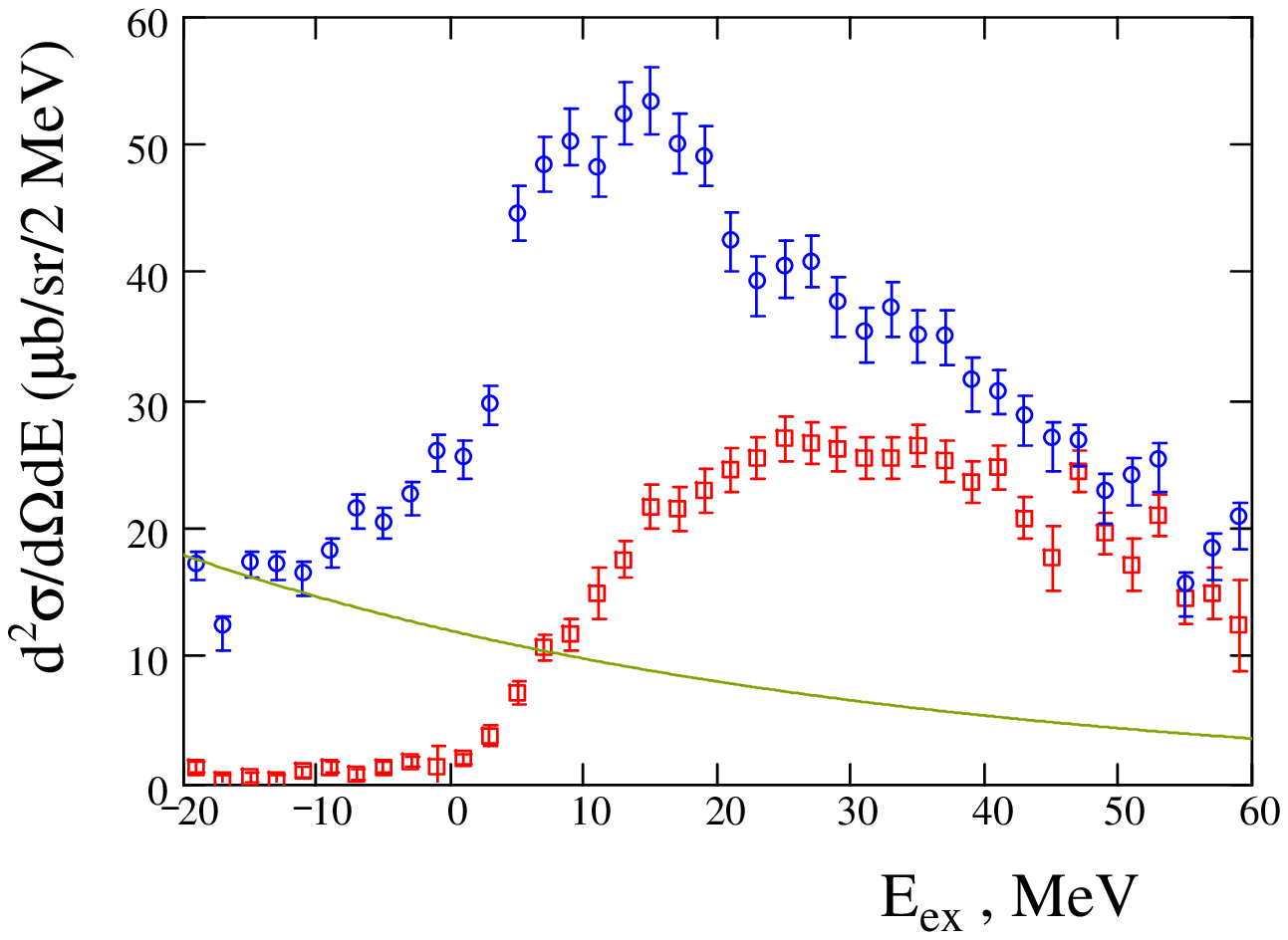, bbllx=90pt, bblly=325pt,
bburx=470pt, bbury=605pt}}
\end{picture}
\caption{The data of ref.[1] on the differential cross sections of
the reactions $^9\mbox{Be}(\mbox{K}^-,{\pi}^-)$ (circles) and
$^9\mbox{Be}(\mbox{K}^-,{\pi}^+)$ (squares) at small angles at 600 MeV/c.
The solid curve is the approximation of the tail of direct $\Lambda$
production used in Section 3.}
\end{figure}

\begin{figure}
\begin{picture}(500,200)
\put(15,10){\epsfig{file=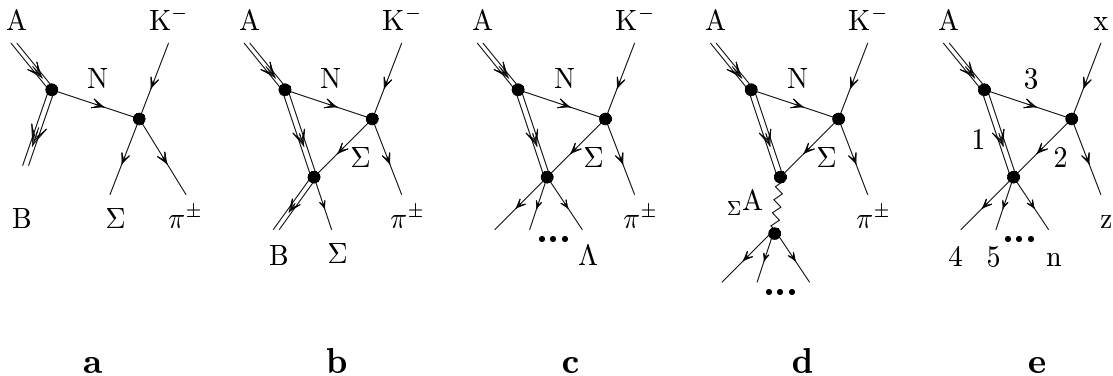, width=420pt}}
\end{picture}
\caption{The graphs for the processes $({\mbox{K}}^-, {\pi}^{\pm})$
on nuclei (a--d) and a generic form of the triangle graph (e).}
\end{figure}

\begin{figure}
\begin{picture}(500,450)
\put(100,10){\epsfig{file=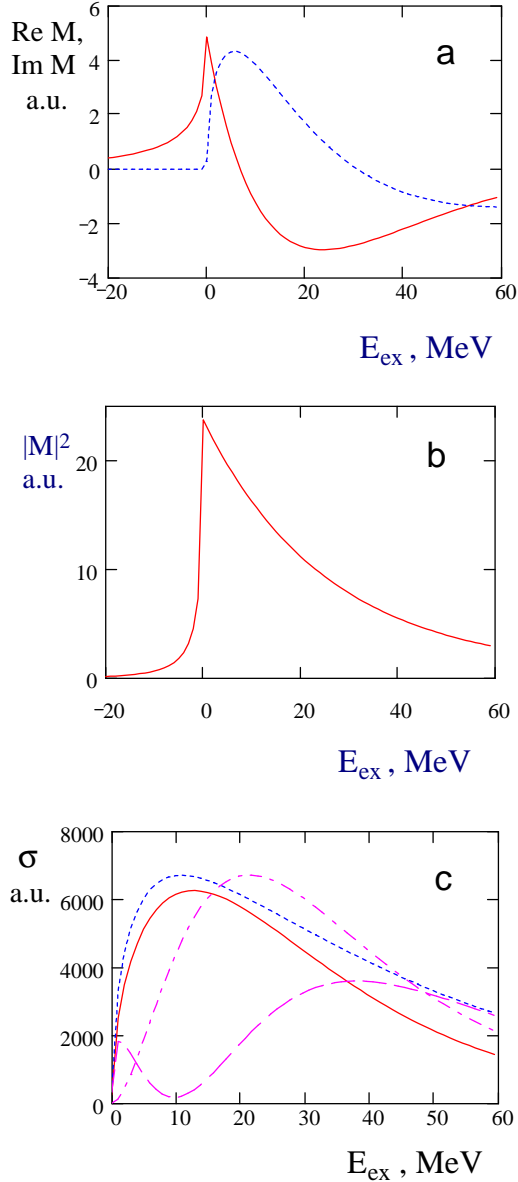, height=450pt, bbllx=125pt, bblly=12pt,
bburx=415pt, bbury=670pt}}
\end{picture}
\caption{The results of intermediate calculations for the process
$^9\mbox{Be}(\mbox{K}^-,{\pi}^+)$: (a) real and imaginary parts of
the triangle graph amplitude; (b) modulus squared of the triangle graph
amplitude; (c) the shapes of the contributions to the cross section from
the quasi--free $\Sigma$ production of Fig.2a (solid curve), from the
Fig.2b triangle graph (dotted curve) and from two versions of the account
of the interference of the Fig.2a and 2b graphs with the relative phase
$0.4{\pi}$ (dashed curve) and $0.9{\pi}$ (dash--dotted curve).}
\end{figure}

\begin{figure}
\begin{picture}(500,450)
\put(70,10){\epsfig{file=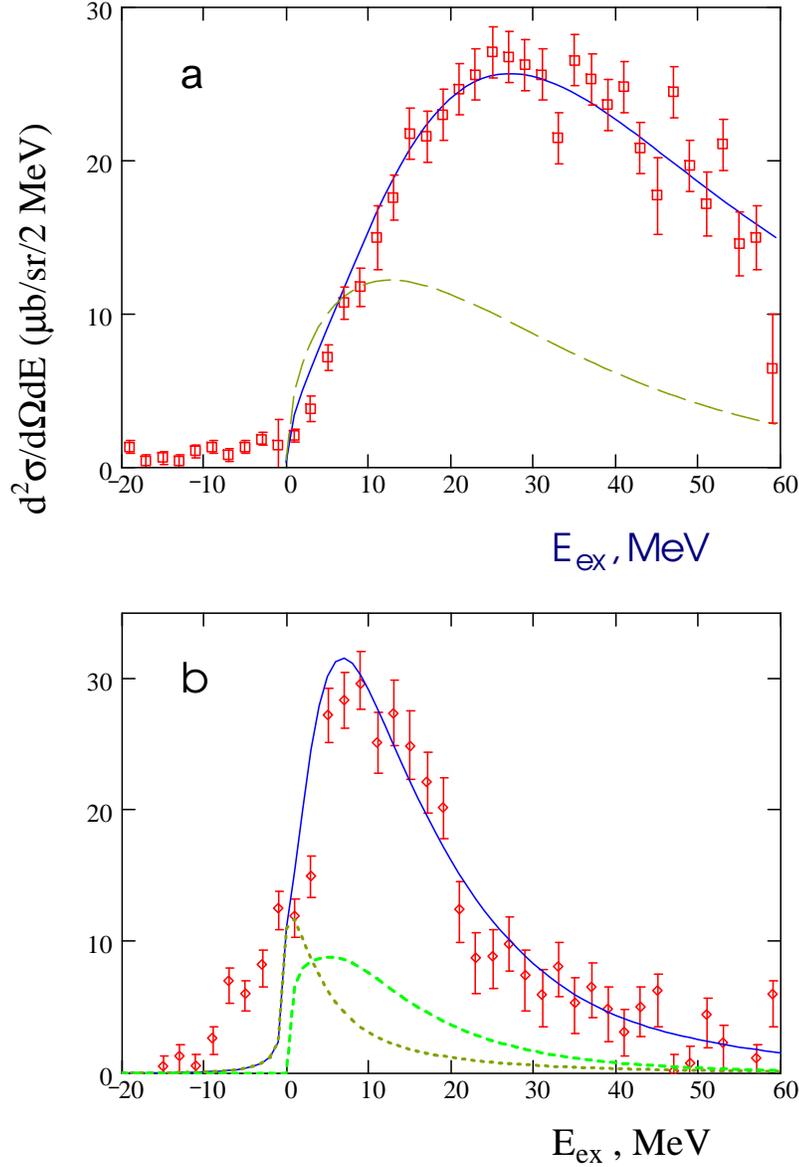, height=450pt, bbllx=90pt, bblly=60pt,
bburx=465pt, bbury=620pt}}
\end{picture}
\caption{(a) The excitation energy distribution in the reaction
$^9\mbox{Be}(\mbox{K}^-,{\pi}^+)$. The data are from ref.[1]. The solid
curve is the result of a full calculation. The dashed curve shows the
contribution only from the quasi--free $\Sigma$ production (Fig.2a).
(b) The excitation energy distribution in the reaction
$^9\mbox{Be}(\mbox{K}^-,{\pi}^-)$ on the outer neutron. The experimental
data are obtained from the data of ref.[1] by means of the difference
procedure described in Section 3. The solid curve is the result of a full
calculation. The dotted curve is the contribution of the processes with
the conversion $\Sigma \to \Lambda$. The dashed curve shows the
contribution only from the quasi--free $\Sigma$ production (Fig.2a).}
\end{figure}

\begin{figure}
\begin{picture}(500,450)
\put(70,10){\epsfig{file=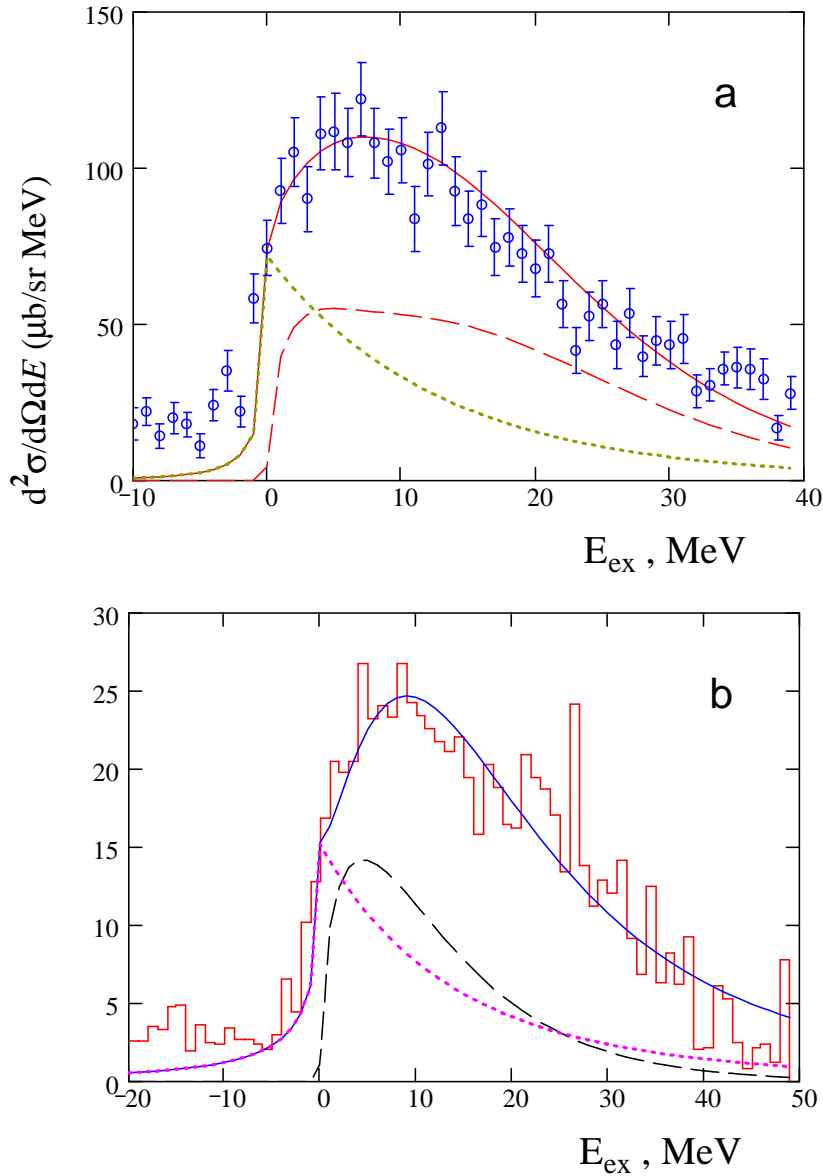, height=450pt, bbllx=90pt, bblly=65pt,
bburx=475pt, bbury=615pt}}
\end{picture}
\caption{(a) The excitation energy distribution in the reaction
$^{12}\mbox{C}(\mbox{K}^-,{\pi}^+)$ at 715 MeV/c for $4^{\circ}$.
The data are from ref.[12]. The solid
curve is the result of a full calculation. The dashed curve shows the
contribution only from the quasi--free $\Sigma$ production (Fig.2a).
(b) The excitation energy distribution in the reaction
$^4\mbox{He}(\mbox{K}^-,{\pi}^+)$ at 600 MeV/c for small angles. The
experimental histogram is from ref.[11]. The meaning of the curves is the
same as in Fig.5a.}
\end{figure}

\begin{figure}
\begin{picture}(500,250)
\put(50,10){\epsfig{file=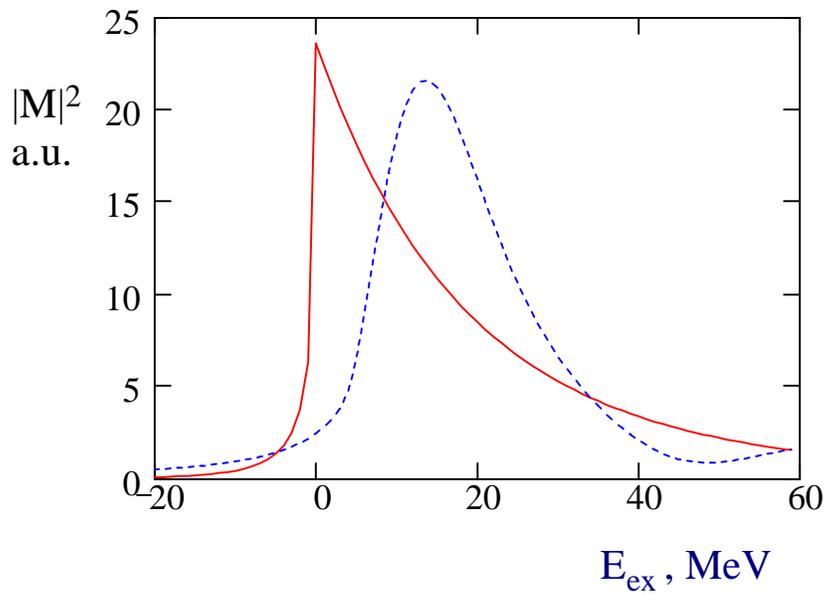, bbllx=145pt, bblly=493pt,
bburx=460pt, bbury=725pt}}
\end{picture}
\caption{$|M|^2$ for the triangle graph of Fig.2c with two--particle
(solid curve) and three--particle (dotted curve) intermediate states.}
\end{figure}

\begin{figure}
\begin{picture}(500,300)
\put(50,10){\epsfig{file=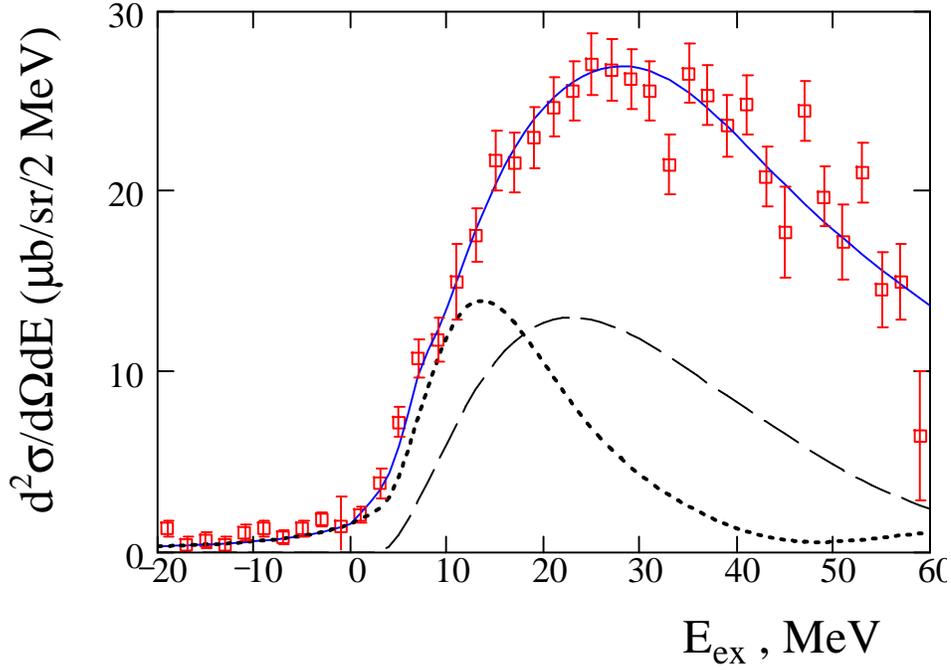, bbllx=90pt, bblly=330pt,
bburx=455pt, bbury=590pt}}
\end{picture}
\caption{The excitation energy distribution in the reaction
$^9\mbox{Be}(\mbox{K}^-,{\pi}^+)$ with three--particle intermediate
state in Fig.2c graph. The solid curve is the result of a full
calculation. The dotted curve is the contribution of the processes with
the conversion $\Sigma \to \Lambda$. The dashed curve shows the
contribution only from the quasi--free $\Sigma$ production.}
\end{figure}

\begin{figure}
\begin{picture}(500,450)
\put(70,10){\epsfig{file=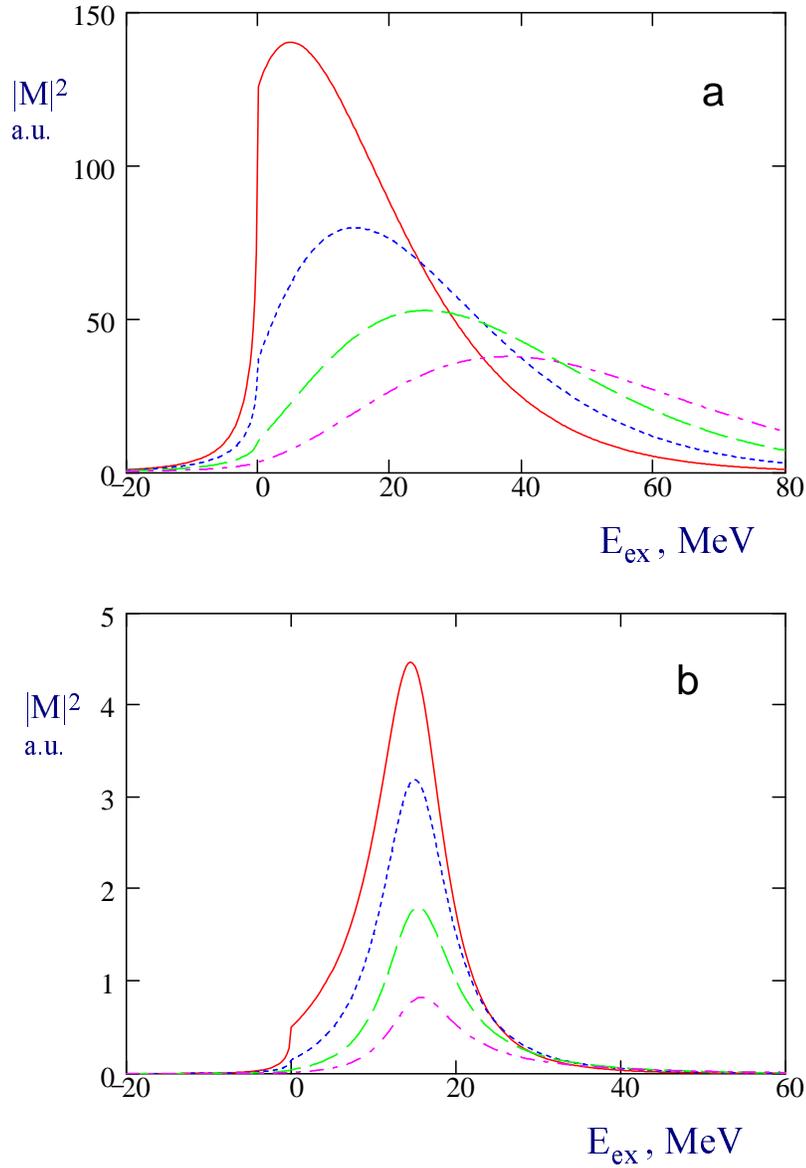, height=450pt, bbllx=110pt, bblly=140pt,
bburx=492pt, bbury=700pt}}
\end{picture}
\caption{(a) The modulus squared of the Fig.2c graph amplitude for
the reaction $^{12}\mbox{C}(\mbox{K}^-,{\pi}^+)$ at different values of
the momentum transfer $q=200$ MeV/c (solid curve), 250 MeV/c (dotted curve),
300 MeV/c (dashed curve) and 350 MeV/c (dash--dotted curve).
(b) The same with the inclusion of the excited $\Sigma$--hypernuclear
state (Fig.2d) with the width 10 MeV and the mass corresponding to
$E_{\mbox{ex}}=15$ MeV.}
\end{figure}

\end{document}